\documentclass[12pt]{article}

\usepackage{scicite}

\usepackage{graphicx}
\usepackage{lineno}
\usepackage{color}

\newcommand{\hla}{$\mathrm{^{3}_{\Lambda}H}$}
\newcommand{\hala}{$\mathrm{^{3}_{\bar{\Lambda}}\overline{{H}}}$}

\newcommand{\he}{$\mathrm{^{3}He}$}
\newcommand{\ahe}{$\mathrm{^{3}\overline{{He}}}$}
\newcommand{\het}{$\mathrm{^{3}H}$}

\newcommand{\la}{$\mathrm{\Lambda}$}

\usepackage{times}
\newenvironment{sciabstract}{%
\begin{quote} \bf}
{\end{quote}}

\newcounter{lastnote}
\newenvironment{scilastnote}{%
\setcounter{lastnote}{\value{enumiv}}%
\addtocounter{lastnote}{+1}%
\begin{list}%
{\arabic{lastnote}.}
{\setlength{\leftmargin}{.22in}}
{\setlength{\labelsep}{.5em}}}
{\end{list}}

\title{Observation of an Antimatter Hypernucleus}

\author{The STAR Collaboration}

\date{}

\begin{document}

% Double-space the manuscript.

\baselineskip24pt

% Make the title.

\maketitle

% Place your abstract within the special {sciabstract} environment.

\begin{sciabstract}
Nuclear collisions recreate conditions in the universe
microseconds after the Big Bang. Only a very small fraction of the
emitted fragments are light nuclei, but these states are of
fundamental interest. We report the observation of
antihypertritons - composed of an antiproton, antineutron, and
antilambda hyperon - produced by colliding gold nuclei at high
energy. Our analysis yields 70 $\pm$ 17 antihypertritons (\hala)
and 157 $\pm$ 30 hypertritons (\hla). The measured yields of
\hla~(\hala) and \he~(\ahe) are similar, suggesting an equilibrium
in coordinate and momentum space populations of up, down, and
strange quarks and antiquarks, unlike the pattern observed at
lower collision energies. The production and properties of
antinuclei, and nuclei containing strange quarks, have
implications spanning nuclear/particle physics, astrophysics, and
cosmology.
\end{sciabstract}

%------------------------------------
%\section*{Introduction}
%------------------------------------

Nuclei are abundant in the universe, but antinuclei that are
heavier than the antiproton have been observed only as products of
interactions at particle accelerators~\cite{AMSantiHe3,antid}.
Collisions of heavy nuclei at the Relativistic Heavy-Ion Collider
(RHIC) at Brookhaven National Laboratory (BNL) briefly produce hot
and dense matter that has been interpreted as a quark gluon plasma
(QGP)~\cite{RHIC-whitepapers-2,QGP-PBM} with an energy density
similar to that of the universe a few microseconds after the Big
Bang. This plasma contains roughly equal numbers of quarks and
antiquarks. As a result of the high energy density of the QGP
phase, many strange-antistrange ($s \overline{s}$) quark pairs are
liberated from the quantum vacuum. The plasma cools and
transitions into a hadron gas, producing nucleons, hyperons,
mesons, and their antiparticles.

Nucleons (protons and neutrons) contain only up and down valence
quarks, while hyperons ($\Lambda, \Sigma, \Xi, \Omega$) contain at
least one strange quark in its 3-quark valence set. A hypernucleus
is a nucleus that contains at least one hyperon in addition to
nucleons. All hyperons are unstable, even when bound in nuclei.
The lightest bound hypernucleus is the hypertriton (\hla), which
consists of a \la~hyperon, a proton, and a neutron. The first
observation of any hypernucleus was made in 1952 using a nuclear
emulsion cosmic ray detector~\cite{hyperT1953}. Here, we present
the observation of an antimatter hypernucleus.

\textbf{Production of antinuclei:} Models of heavy-ion collisions
have had good success in explaining the production of nuclei by
assuming that a statistical coalescence mechanism is in effect
during the late stage of the collision
evolution~\cite{QGP-PBM,Sato-Coalescence1981}. Antinuclei can be
produced through the same coalescence mechanism, and are predicted
to be present in cosmic rays. An observed high yield could be
interpreted as an indirect signature of new physics, such as Dark
Matter~\cite{AMSCol,antid2008}. Heavy-ion collisions at RHIC
provide an opportunity for the discovery and study of many
antinuclei and antihypernuclei.

The ability to produce antihypernuclei allows the study of all
populated regions in the 3-dimensional chart of the nuclides. The
conventional 2-dimensional chart of the nuclides organizes nuclear
isotopes in the ($N$, $Z$) plane, where $N$ is the number of
neutrons and the $Z$ is the number of protons in the nucleus. This
chart can be extended to the negative sector in the ($N$, $Z$)
plane by including antimatter nuclei. Hypernuclei bring a third
dimension into play, based on the strangeness quantum number of
the nucleus. The present study probes the territory of antinuclei
with non-zero strangeness (Fig. 1), where proposed ideas
\cite{Greiner95,QGP_r21,QGP_r22,QGP_r23} related to the structure
of nuclear matter can be explored.

\textbf{Hypernuclei-Formation and observation:} The
hyperon-nucleon (YN) interaction, responsible in part for the
binding of hypernuclei, is of fundamental interest in nuclear
physics and nuclear astrophysics. For example, the YN interaction
plays an important role in attempts to understand the structure of
neutron stars. Depending on the strength of the YN interaction,
the collapsed stellar core could be composed of hyperons, of
strange quark matter, or of a kaon condensate~\cite{neuteronstar}.
While the hyperons or strange particles inside a dense neutron
star would not decay because of local energy constraints, free
hypernuclei decay into ordinary nuclei with typical lifetimes of a
few hundred picoseconds, which is still thirteen orders of
magnitude longer than the lifetimes of the shortest-lived
particles. The lifetime of a hypernucleus depends on the strength
of the YN interaction~\cite{Dalitz1962,Glockle1998}. Therefore, a
precise determination of the lifetime of hypernuclei provides
direct information on the YN interaction
strength~\cite{hyperTBE1973,Glockle1998}.
\begin{figure}
\includegraphics
[scale=0.45,bb=80 50 120 560]{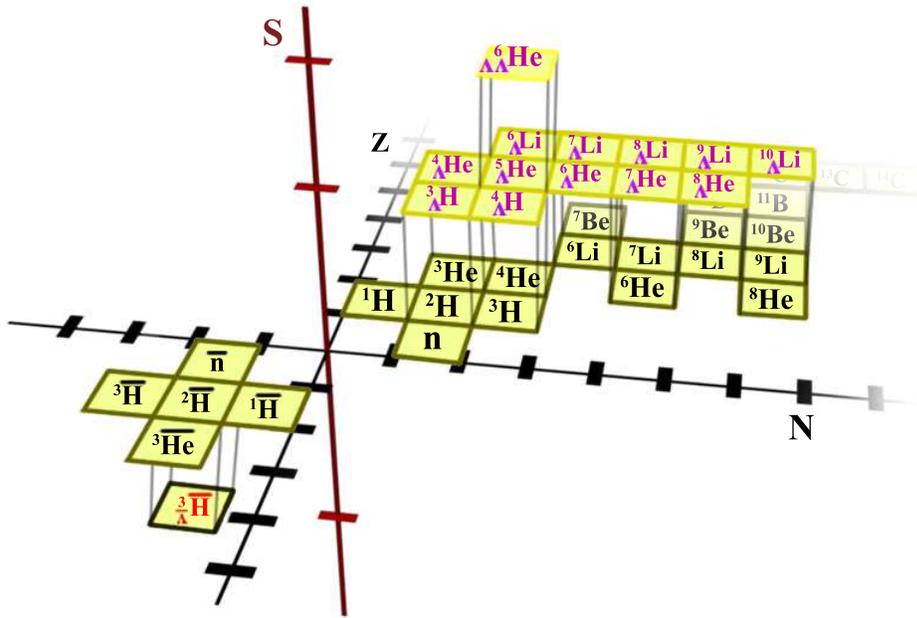} \caption{A
chart of the nuclides showing the extension into the strangeness
sector. Normal nuclei lie in the ($N$, $Z$) plane. Antinuclei lie
in the negative sector of this plane. Normal hypernuclei lie in
the positive ($N$, $Z$) quadrant above the plane. The
antihypertriton \hala~reported here extends this chart into the
strangeness octant below the antimatter region in the ($N$, $Z$)
plane.}
\end{figure}

%--------------------------------------
%\section*{Experiment and Analysis}
%--------------------------------------

The experiment was carried out by the STAR
collaboration~\cite{STAR2003} at the RHIC facility. The main
detector of the STAR experiment is a gas-filled cylindrical Time
Projection Chamber (TPC), with an inner radius of 50 cm, an outer
radius of 200 cm, and a length of 420 cm along the beam
line~\cite{TPC2003}. The TPC is a device for imaging, in three
dimensions, the ionization left along the path of charged
particles. It resolves over 50 million pixels within its active
volume. The present analysis is based on interactions produced by
colliding two Au beams at an energy of 200 GeV per nucleon-nucleon
collision in the center-of-mass system. Approximately 89 million
collision events were collected using a trigger designed to
accept, as far as possible, all impact parameters (minimum-bias
event), and an additional 22 million events were collected using a
trigger that preferentially selects near-zero impact parameter (or
``head-on") collisions. The accepted collisions are required to
occur within 30 cm of the center of the TPC along the beam line.
Charged particle tracks traversing the TPC are reconstructed in an
acceptance that is uniform in azimuthal angle. The precise
coverage in terms of polar angle is somewhat
complicated~\cite{TPC2003}, but roughly speaking, charged tracks
emerging at angles with respect to the beam axis in the range of
$45^\circ < \theta < 135^\circ$ are reconstructed.

\begin{figure}
\includegraphics[scale=0.85,bb=30 -20 10 200]{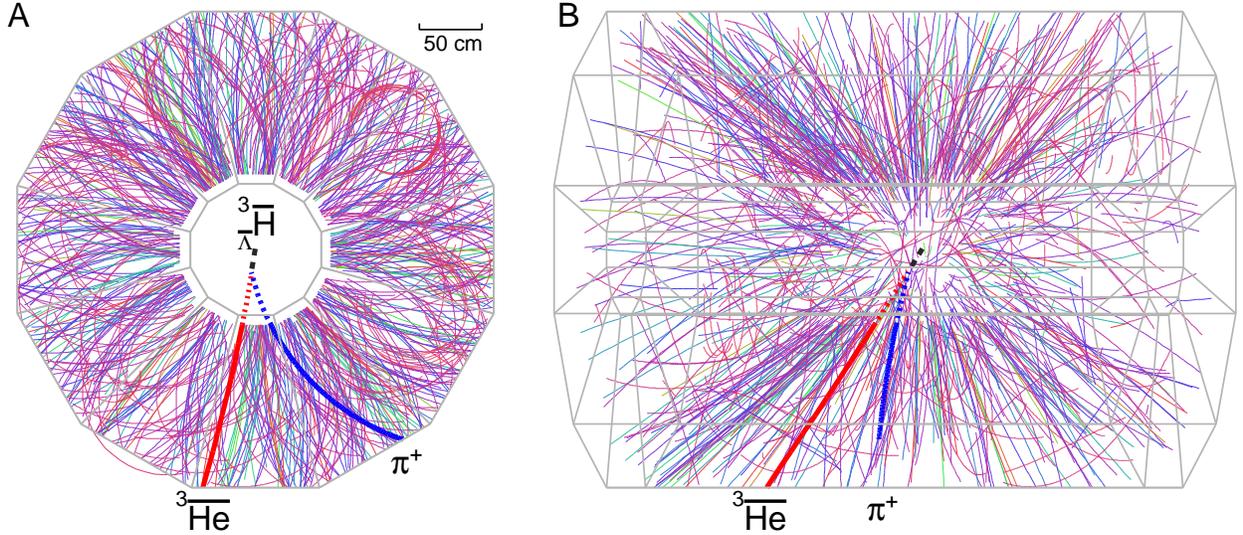}
\caption{A typical event in the STAR detector that includes the
production and decay of a \hala~candidate. In (A), the beam axis
is normal to the page, and in (B), the beam axis is horizontal.
The dashed black line is the trajectory of the \hala~candidate,
which cannot be directly measured. The heavy red and blue lines
are the trajectories of the \ahe~and $\mathrm{\pi^{+}}$ decay
daughters, respectively, which are directly measured.}
\end{figure}

Fig. 2 depicts a typical Au$+$Au collision reconstructed in the
STAR TPC.  The tracks are curved by a uniform magnetic field of
0.5 Tesla parallel to the beam line. The event of interest here
includes a \hala~candidate created at the primary collision vertex
near the center of the TPC. The \hala~travels a few centimeters
before it decays. One of the possible decay channels is \hala
$\rightarrow$ \ahe $+\pi^+$, which occurs with a branching ratio
of 25\% assuming that this branching fraction is the same as that
for \hla~\cite{Glockle1998}. The two daughter particles then
traverse the TPC along with the hundreds of other charged
particles produced in the primary Au$+$Au collision. The
trajectories of the daughter particles are reconstructed from the
ionization trails they leave in the TPC gas volume (shown in Fig.
2 as thick red and blue lines for \ahe~and $\pi^{+}$,
respectively). The energy loss by these particles to ionization in
the TPC, $\langle dE/dx \rangle$, depends on the particle velocity
and charge. Particle identification is achieved by correlating the
$\langle dE/dx \rangle$ values for charged particles in the TPC
with their measured magnetic rigidity, which is proportional to
the inverse of the curvature of the trajectory in the magnetic
field. With both daughter candidates directly identified, one can
trace back along the two helical trajectories to the secondary
decay point, and thereby reconstruct the location of the decay
vertex as well as the parent momentum vector.

\textbf{Particle identification:} Fig. 3 presents results from the
antihypertriton analysis outlined above, along with results from
applying the same analysis to measure the normal matter
hypertritons in the same dataset --- only the sign of the
curvature of the decay products is reversed. Fig. 3C shows
$\langle dE/dx \rangle$ for negative tracks as a function of the
magnetic rigidity; the different bands result from the different
particle species. The measured $\langle dE/dx \rangle$ of the
particles is compared to the expected value from the Bichsel
function~\cite{PDG2008}, which is an extension of the usual Bethe
Bloch formulas for energy loss. A new variable, $z$, is defined as
$ z= \ln(\langle dE/dx \rangle / \langle dE/dx \rangle_B)$, where
$\langle dE/dx \rangle_B$ is the expected value of $\langle dE/dx
\rangle$ for the given particle species and momentum. The measured
$z({\mathrm{^{3}He}})$ distributions for \he~and \ahe~tracks (Fig.
3D), includes 5810 \he~and 2168 \ahe~candidates with $|z(
\mathrm{^3He})| < 0.2$, and represents the largest sample of
\ahe~antinuclei that has been collected to date. The first few
\ahe~candidates were observed at the Serpukhov accelerator
laboratory~\cite{antiHe31971}, followed by confirmation from the
European Organization for Nuclear Research
(CERN)~\cite{NA52antiHe3}. In 2001, a relatively large \ahe~sample
was reported by the STAR collaboration~\cite{STARantiHe3}. The
\he~and \ahe~samples in the present analysis are so cleanly
identified that misidentification from other weak decays is
negligible. However, due to the $\langle dE/dx \rangle$ overlap
between \het~and \he~at low momenta, it is only possible to
identify the \he~nuclei at relatively high momenta ({\it i.e.}
above $\sim$$2$~GeV/$c$). The daughter pions from \hala~decays
usually have momenta $\sim$ 0.3 GeV/$c$, and can be cleanly
identified~\cite{PID2006}.

%-------------------------------------
%\section*{Results: Invariant Mass}
%-------------------------------------

\textbf{Topological reconstruction:} A set of topological cuts is
invoked in order to identify and reconstruct the secondary decay
vertex positions with a high signal-to-background ratio. These
cuts involve the distance at the decay vertex between the tracks
for the \ahe~and $\pi^+$ ($<$1 cm), the distance of closest
approach (DCA) between the \hala~candidate and the event primary
vertex ($<$1 cm), the decay length of the \hala~candidate vertex
from the event primary vertex ($>$2.4 cm), and the DCA between the
$\pi$ track and the event primary vertex ($>$0.8 cm). The cuts are
optimized based on full detector response
simulations~\cite{hyperon2007}. Several different cut criteria are
also applied to cross-check the results and to estimate the
systematic errors. The signal is always present, and the
difference in the total yields using different cuts are found to
be less than 15\%. The total systematic error in the present
analysis is 15\%.

The parent candidate invariant mass is calculated based on the
momenta of the daughter candidates at the decay vertex. The
results are shown as the open circles in Fig. 3A for the
hypertriton: $\mathrm{^{3}_{\Lambda}H \rightarrow ^{3}He +
\pi^{-}}$, and in Fig. 3B for the antihypertriton:
\hala~$\rightarrow$~\ahe $ +\pi^{+}$. There remains an appreciable
combinatorial background in this analysis, which must be described
and subtracted. A track rotation method is used to reproduce this
background. This approach involves the azimuthal rotation of the
daughter \he~(\ahe) track candidates by 180 degrees with
\begin{figure}
\includegraphics[scale=0.75,bb=-20 -30 50 340]{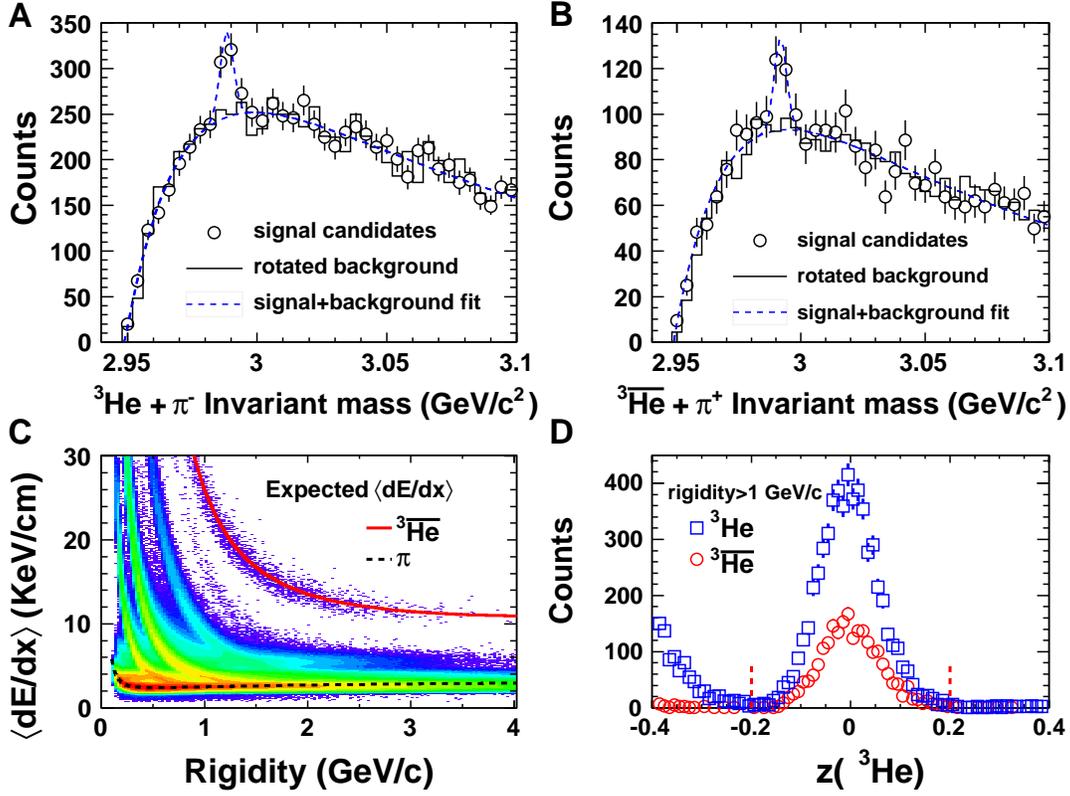}
\vspace{-0.55cm}\caption{(A, B) show the invariant mass
distribution of the daughter $\mathrm{^{3}He + \pi}$. The open
circles represent the signal candidate distributions, while the
solid black lines are background distributions. The blue dashed
lines are signal (Gaussian) plus background (double exponential)
combined fit (see the text for details). A (B) shows the
\hla~(\hala) candidate distributions. (C) shows $\langle dE/dx
\rangle$ versus rigidity
($\mathrm{momentum/|nuclear~charge~units|}$) for negative tracks.
Also plotted are the expected values for \ahe~and $\pi$ tracks.
(D) and (C) demonstrate that the \he~and \ahe~tracks ($|z(^{3}{\rm
He})|<0.2$) are identified essentially without background.}
\end{figure}
respect to the event primary vertex. In this way, the event is not
changed statistically, but all of the secondary decay topologies
are destroyed because one of the daughter tracks is rotated away.
This provides an accurate description of the combinatorial
background. The resulting rotated invariant mass distribution is
consistent with the background distribution, as shown by the solid
histograms (Fig. 3A,B). The rotated background distribution is fit
with a double exponential function: $f(x)\propto
\exp(-\frac{x}{p_1}) - \exp(-\frac{x}{p_2})$, where $x=m-m(^3{\rm
He})-m(\pi)$, and $p_1,~p_2$ are fit parameters. Finally, the
counts in the signal are calculated after subtraction of this fit
function derived from the rotated background. In total, $157 \pm
30$ \hla~and 70 $\pm$ 17 \hala~candidates are thus observed. The
quoted errors are statistical.

\textbf{Production and properties:} We can use the measured
\hla~yield to estimate the expected yield of the \hala, assuming
symmetry between matter and antimatter, in the following manner:
\hala~$=$~\hla $ \times $\ahe ~/ \he $= 59 \pm 11$. This indicates
a $5.2\sigma$ projection of the number of \hala~that is expected
in the same data set where \hla, \he~and \ahe~are detected. An
additional check involves fitting the \he~$ + \pi$ invariant mass
distribution with the combination of a Gaussian ``signal'' term
plus the double-exponential background function (the blue-dashed
lines in Fig. 3A,B). The resulting mean values and widths of the
invariant mass distributions are consistent with the results from
the full detector response simulations. Our best fit values (from
$\chi^2$ minimization) are $m$(\hla) $= 2.989 \pm 0.001 \pm 0.002$
GeV$/c^2$ and $m$(\hala) $= 2.991 \pm 0.001 \pm 0.002$ GeV$/c^2$.
These values are consistent with each other within the current
statistical and systematic errors, and are consistent with the
best value from the literature, i.e., $m$(\hla) $= 2.99131 \pm
0.00005$ GeV$/c^2$~\cite{hyperTBE1973}. Our systematic error of 2
MeV$/c^2$ arises from well-understood instrumental effects that
cause small deviations from ideal helical ionization tracks in the
TPC.

%---------------------------------------
%\section*{Results: lifetime}
%---------------------------------------

\begin{figure}
\includegraphics[scale=0.750,bb=-20 -35 100 240]{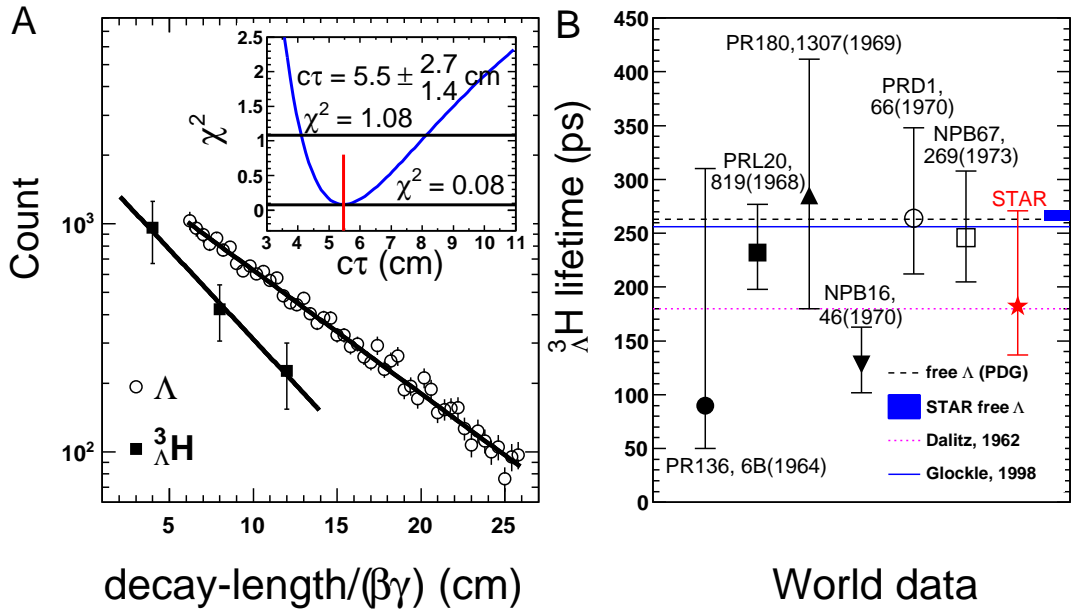}
\vspace{-0.55cm}\caption{(A) The \hla~(solid squares) and
$\mathrm{\Lambda}$ (open circles) yield distributions versus
$c\tau$. The solid lines represent the $c\tau$ fits. The inset
depicts the $\chi^{2}$ distribution of the best \hla~$c\tau$ fit.
(B) World data for \hla~lifetime measurements. The data points are
from
Refs.~\cite{lifetime1964,lifetime1968,lifetime1969,lifetime1970,lifetime19702,lifetime1973}.
The theoretical calculations are from
Refs.~\cite{Dalitz1962,Glockle1998}. The error bars represent the
statistical uncertainties only.}
\end{figure}

\textbf{Lifetimes:} The direct reconstruction of the secondary
decay vertex in this data allows measurement of the \hla~lifetime,
$\tau$, via the equation $N(t)=N(0)e^{-t/\tau},~t=l/(\beta \gamma
c),~ \beta \gamma c=p/m$, where $l$ is the measured decay
distance, $p$ is the particle momentum, $m$ is the particle mass,
and $c$ is the speed of light. For better statistics in our fit,
the \hla~and \hala~samples are combined, as the matter-antimatter
symmetry requires their lifetimes to be equal. Separate
measurements of the lifetimes for the two samples show no
difference within errors. The signal is then plotted in three bins
in $l/\beta\gamma$. The yield in each bin is corrected for the
experimental tracking efficiency and acceptance. The total
reconstruction efficiency of the \hala~and \hla~is on the order of
10\%, considering all sources of loss and the analysis cuts. The
three points are then fit with the exponential function to extract
the parameter $c\tau$, and the best-fit result is displayed as the
solid line in Fig. 4A. To arrive at the optimum fit, a $\chi^2$
analysis was performed (see the inset to Fig. 4A). The $c\tau$
parameter that is observed in this analysis is $c\tau = 5.5 \pm
^{2.7}_{1.4} \pm 0.8$ cm, which corresponds to a lifetime $\tau$
of $182 \pm ^{89}_{45} \pm 27$ ps. As an additional cross-check,
the \la~hyperon lifetime is extracted from the same data set using
the same approach, for the $\mathrm{\Lambda \rightarrow p +
\pi^{-}}$ decay channel. The result obtained is $\tau = 267 \pm 5
(stat)$ ps, which is consistent with $\tau=263 \pm 2$ ps compiled
by the Particle Data Group~\cite{PDG2008}.

The \hla~lifetime measurements to
date~\cite{lifetime1963,lifetime1964,lifetime1970,lifetime1968,lifetime1969,lifetime19702,lifetime1973}
are not sufficiently accurate to distinguish between models, as
depicted by Fig. 4B. The present measurement is consistent with a
calculation using a phenomenological \hla~wave
function~\cite{Dalitz1962}, and is also consistent with a more
recent three-body calculation~\cite{Glockle1998} using a more
modern description of the baryon-baryon force. The present result
is also comparable to the lifetime of free $\Lambda$ particles
within the uncertainties, and is statistically competitive with
the earlier experimental measurements.

%------------------------------------------------
%\section*{Results: particle ratio, correlation}
%------------------------------------------------

\begin{table}[h!]
\centering \caption{ Particle ratios from Au+Au collisions at %%% $\sqrt{s_{NN}}$ =
200 GeV. } \label{table1}
%\begin{ruledtabular}
\begin{tabular}
{cc}\\ \hline \hline Particle type & Ratio\\\hline
\\
\hala/\hla &  0.49 $\pm$ 0.18 $\pm$ 0.07 \\
\\
\ahe/\he   &  0.45 $\pm$ 0.02 $\pm$ 0.04 \\
\\
\hala/\ahe &  0.89 $\pm$ 0.28 $\pm$ 0.13 \\
\\
\hla/\he   &  0.82 $\pm$ 0.16 $\pm$ 0.12 \\\hline\hline
\end{tabular}
%\end{ruledtabular}
\end{table}

\textbf{Coalescence calculations:} The coalescence model makes
specific predictions about the ratios of particle yields. These
predictions can be checked for a variety of particle species. To
determine the invariant particle yields of \hala~and \hla,
corrections for detector acceptance and inefficiency are applied.
The \hala~and \hla~yields are measured in three different
transverse momentum ($p_t$) bins within the analyzed transverse
momentum region of $2 < p_t < 6$ GeV/$c$ and then extrapolated to
the unmeasured regions ($p_t<2$ GeV/$c$ and $p_t>6$ GeV/$c$). This
extrapolation assumes that both \hala~and \hla~have the same
spectral shape as the high-statistics \ahe~and \he~samples from
the same data set (see Table 1). If the \hala~and \hla~are formed
by coalescence of ($\bar{\Lambda} + \bar{p} + \bar{n}$) and
($\Lambda + p + n$), then the production ratio of \hala~to
\hla~should be proportional to ($\frac{\bar{\Lambda}}{\Lambda}
\times \frac{\bar{p}}{p} \times \frac{\bar{n}}{n}$). The latter
value can be extracted from spectra already measured by STAR, and
the value obtained is $0.45 \pm 0.08 \pm
0.10$~\cite{PID2006,hyperon2007}. The measured \hala~/~\hla~and
\ahe~/~\he~ratios are consistent with the interpretation that the
\hala~and \hla~are formed by coalescence of ($\bar{\Lambda} +
\bar{p} + \bar{n}$) and ($\Lambda + p + n$), respectively.

\textbf{Discussion:} As the coalescence process for the formation
of (anti)hypernuclei requires that (anti)nucleons and
(anti)hyperons be in proximity in phase space (i.e., in coordinate
and momentum space), (anti)hypernucleus production is sensitive to
the correlations in phase-space distributions of nucleons and
hyperons \cite{Sato-Coalescence1981}. An earlier two-particle
correlation measurement published by STAR implies a strong
phase-space correlation between protons and \la~hyperons
\cite{STAR-p-la}. Equilibration among the strange quark flavors
and light quark flavors is one of the proposed signatures of QGP
formation~\cite{Rafelski-PRL1982}, which would result in high
(anti)hypernucleus yields. In addition, recent theoretical studies
motivate a search for the onset of QGP by studying the evolution
of the baryon -- strangeness correlation as a function of
collision energy~\cite{Cbs_PRL,Cbs_PRC,Cbs_LQCD}. The \hla~yields
provide a natural and sensitive tool to extract this
correlation~\cite{Sato-Coalescence1981,ampt}, as they can be
compared to the yields of \he~and \het, which have the same atomic
mass number. Besides $4u+4d$ valence quarks, the valence quark
content of these species includes one additional $u,~d$ and $s$
quark for \he, \het~and \hla, respectively. Recent nuclear
transport model calculations~\cite{ampt} support the expectation
that the strangeness population factor%%~\cite{ampt}
,$S_3=$\hla/(\he~$\times \Lambda/p$), can be used as a tool to
distinguish the QGP from a purely hadronic phase.

\begin{figure}[htbp]
\includegraphics[scale=0.50,bb=-180 -20 100 520]{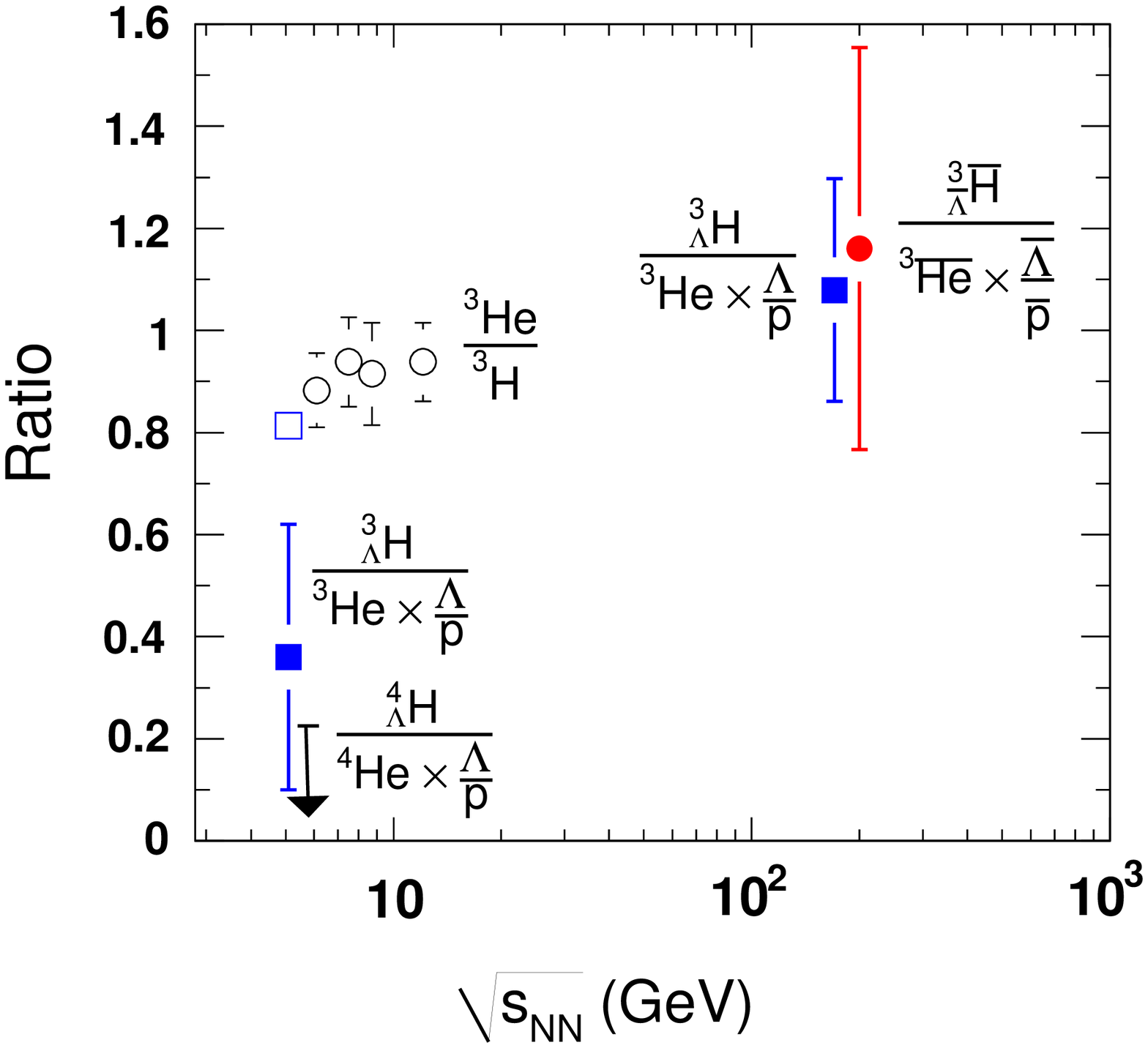}
\vspace{-0.55cm}\caption{Particle ratios as a function of
center-of-mass energy per nucleon-nucleon collision. The data at lower
energies are from Refs.~\cite{AGS2004,AGS2000,SPS2008}. The
$\mathrm{^{4}_{\Lambda}H/(^{4}He \times \Lambda/p)}$ ratio is
corrected for the spin degeneracy factor~\cite{AGS2004}. The error
bars represent statistical uncertainties only.}
\end{figure}

Fig. 5 depicts various particle ratios as a function of the
collision energy. The \he/\het~ratio at a center-of-mass energy of
5 GeV obtained at the Alternating Gradient Synchrotron (AGS) at
BNL is much closer to unity than the ratio \hla/\he~at the same
energy. The values of $S_3$ are about 1/3 at AGS energies, and
near unity at RHIC energies, although with large uncertainties.
The AGS value is further constrained to be relatively low by the
measured upper limit on the $\mathrm{^{4}_{\Lambda}H/^{4}He}$
ratio~\cite{AGS2004}, indicating that the phase space population
for strangeness is very similar to that for the light quarks in
high-energy heavy-ion collisions at RHIC, in contrast to the
situation at AGS.

Individual relativistic heavy-ion collisions produce abundant
hyperons containing one ($\Lambda,~\Sigma$), two ($\Xi$) or three
($\Omega$) strange (anti)quarks. The coalescence mechanism for
hypernucleus production in these collisions thus provides a source
for other exotic hypernucleus searches. This should allow an
extension of the 3-D chart of the nuclides (Fig. 1) further into
the antimatter sectors. Future RHIC running will provide increased
statistics, allowing detailed studies of masses and lifetimes, as
well as stringent tests of production rates compared to
predictions based on coalescence models.

%---------------------------------------
%\section*{Concluding remarks}
%---------------------------------------
\textbf{Concluding remark:} Evidence for the observation of an
antihypernucleus, the \hala, with a statistical significance of
$4.1\sigma$ has been presented; consistency checks and constraints
from a \hla~analysis in the same event sample, with $5.2\sigma$
significance, support this conclusion. The lifetime is observed to
be $\tau = 182 \pm ^{89}_{45} \pm 27$ ps, which is comparable to
that of the free $\Lambda$ hyperon within current uncertainties.
The \hala~(\hla) to \ahe~(\he) ratio is close to unity and is
significantly larger than that measured at lower beam energies,
indicating that the strangeness phase space population is similar
to that of light quarks. An order-of-magnitude larger sample of
similar collisions is scheduled to be recorded in the near future.
The antihypernucleus observation demonstrates that RHIC is an
ideal facility for producing exotic hypernuclei and antinuclei.

%---------------------------------------
%\section*{Reference}
%---------------------------------------
\bibliography{scibib}

\begin{thebibliography}{10}

\bibitem{AMSantiHe3}
J.~Alcaraz {\it et~al.\/}, {\it Phys. Lett. B\/} {\bf 461}, 387 (1999).

\bibitem{antid}
H.~Fuke {\it et~al.\/}, {\it Phys. Rev. Lett.\/} {\bf 95}, 081101 (2005).

\bibitem{RHIC-whitepapers-2}
J.~Adams {\it et~al.\/}, {\it Nucl. Phys. A\/} {\bf 757}, 102 (2005).

\bibitem{QGP-PBM}
P.~Braun-Munzinger, J.~Stachel, {\it Nature\/} {\bf 448}, 302 (2007). Also
  references therein.

\bibitem{hyperT1953}
M.~Danysz, J.~Pniewski, {\it Phil. Mag.\/} {\bf 44}, 348 (1953).

\bibitem{Sato-Coalescence1981}
H.~Sato, K.~Yazaki, {\it Phys. Lett. B\/} {\bf 98}, 153 (1981).

\bibitem{AMSCol}
S.~Ahlen {\it et~al.\/}, {\it Nucl. Instrum. Methods A\/} {\bf 350}, 351
  (1994).

\bibitem{antid2008}
F.~Donato, N.~Fornengo, D.~Maurin, {\it Phys. Rev. D\/} {\bf 78}, 043506
  (2008).

\bibitem{Greiner95}
W.~Greiner, {\it Int. J. Mod. Phys. E\/} {\bf 5}, 1 (1996).

\bibitem{QGP_r21}
U.~Heinz, P.~R. Subramanian, H.~St\"ocker, W.~Greiner, {\it J. Phys. G: Nucl.
  Phys.\/} {\bf 12}, 1237 (1986).

\bibitem{QGP_r22}
C.~Greiner, D.-H. Rischke, H.~St\"ocker, P.~Koch, {\it Phys. Rev. D\/} {\bf
  38}, 2797 (1988).

\bibitem{QGP_r23}
J.~Schaffner, C.~Greiner, H.~St\"ocker, {\it Phys. Rev. C\/} {\bf 46}, 322
  (1992).

\bibitem{neuteronstar}
J.~M. Lattimer, M.~Prakash, {\it Science\/} {\bf 304}, 536 (2004).

\bibitem{Dalitz1962}
R.~H. Dalitz, G.~Rajasekharan, {\it Phys. Lett.\/} {\bf 1}, 58 (1962).

\bibitem{Glockle1998}
H.~Kamada, J.~Golak, K.~Miyagawa, H.~Witala, W.~Gl\"ockle, {\it Phys. Rev. C\/}
  {\bf 57}, 1595 (1998).

\bibitem{hyperTBE1973}
M.~Juric {\it et~al.\/}, {\it Nucl. Phys. B\/} {\bf 52}, 1 (1973).

\bibitem{STAR2003}
K.~H. Ackermann {\it et~al.\/}, {\it Nucl. Instrum. Methods A\/} {\bf 499},
  624 (2003).

\bibitem{TPC2003}
M.~Anderson {\it et~al.\/}, {\it Nucl. Instrum. Methods A\/} {\bf 499}, 659
  (2003).

\bibitem{PDG2008}
C.~Amsler {\it et~al.\/}, {\it Phys. Lett. B\/} {\bf 667}, 1 (2008).

\bibitem{antiHe31971}
Y.~Antipov {\it et~al.\/}, {\it Sov. J. Nucl. Phys.\/} {\bf 12}, 171 (1971).

\bibitem{NA52antiHe3}
G.~Ambrosini {\it et~al.\/}, {\it Heavy Ion Phys.\/} {\bf 14}, 297 (2001).

\bibitem{STARantiHe3}
C.~Adler {\it et~al.\/}, {\it Phys. Rev. Lett.\/} {\bf 87}, 262301 (2001).

\bibitem{PID2006}
B.~I. Abelev {\it et~al.\/}, {\it Phys. Rev. Lett.\/} {\bf 97}, 152301 (2006).

\bibitem{hyperon2007}
J.~Adams {\it et~al.\/}, {\it Phys. Rev. Lett.\/} {\bf 98}, 062301 (2007).

\bibitem{lifetime1964}
R.~J. Prem, P.~H. Steinberg, {\it Phys. Rev.\/} {\bf 136}, B1803 (1964).

\bibitem{lifetime1968}
G.~Keyes {\it et~al.\/}, {\it Phys. Rev. Lett.\/} {\bf 20}, 819 (1968).

\bibitem{lifetime1969}
R.~E. Phillips, J.~Schneps, {\it Phys. Rev.\/} {\bf 180}, 1307 (1969).

\bibitem{lifetime1970}
G.~Bohm {\it et~al.\/}, {\it Nucl. Phys. B\/} {\bf 16}, 46 (1970).

\bibitem{lifetime19702}
G.~Keyes {\it et~al.\/}, {\it Phys. Rev. D\/} {\bf 1}, 66 (1970).

\bibitem{lifetime1973}
G.~Keyes, J.~Sacton, J.~H. Wickens, M.~M. Block, {\it Nucl. Phys. B\/} {\bf
  67}, 269 (1973).

\bibitem{lifetime1963}
M.~M. Block {\it et~al.\/}, {\it {Proceedings of the International Conference
  on Hyperfragments at St. Cergue 1963, CERN report 64-1}\/} p.~63 (1964).

\bibitem{STAR-p-la}
J.~Adams {\it et~al.\/}, {\it Phys. Rev. C\/} {\bf 74}, 064906 (2006).

\bibitem{Rafelski-PRL1982}
J.~Rafelski, B.~M\"uller, {\it Phys. Rev. Lett.\/} {\bf 48}, 1066 (1982).

\bibitem{Cbs_PRL}
V.~Koch, A.~Majumder, J.~Randrup, {\it Phys. Rev. Lett.\/} {\bf 95}, 182301
  (2005).

\bibitem{Cbs_PRC}
A.~Majumder, B.~M\"uller, {\it Phys. Rev. C\/} {\bf 74}, 054901 (2006).

\bibitem{Cbs_LQCD}
R.~V. Gavai, S.~Gupta, {\it Phys. Rev. D\/} {\bf 73}, 014004 (2006).

\bibitem{ampt}
S.~Zhang {\it et~al.\/}, {\it Phys. Lett. B\/} {\bf 684}, 224 (2010).

\bibitem{AGS2004}
T.~A. Armstrong {\it et~al.\/}, {\it Phys. Rev. C\/} {\bf 70}, 024902 (2004).

\bibitem{AGS2000}
T.~A. Armstrong {\it et~al.\/}, {\it Phys. Rev. C\/} {\bf 61}, 064908 (2000).

\bibitem{SPS2008}
{V. I. Kolesnikov for the NA49 Collaboration}, {\it J. Phys. Conf. Ser.\/} {\bf
  110}, 032010 (2008).

\end{thebibliography}
\bibliographystyle{Science}
%---------------------------------------
%\section*{Acknowledgement}
%---------------------------------------
\begin{scilastnote}
\item We thank K. Synder for providing Fig. 1. We thank the RHIC
Operations Group and RCF at BNL, the NERSC Center at LBNL and the
Open Science Grid consortium for providing resources and support.
This work was supported in part by the Offices of NP and HEP
within the U.S. DOE Office of Science, the U.S. NSF, the Sloan
Foundation, the DFG cluster of excellence `Origin and Structure of
the Universe', CNRS/IN2P3, STFC and EPSRC of the United Kingdom,
FAPESP CNPq of Brazil, Ministry of Ed. and Sci. of the Russian
Federation, NNSFC, CAS, MoST, and MoE of China, GA and MSMT of the
Czech Republic, FOM and NOW of the Netherlands, DAE, DST, and CSIR
of India, Polish Ministry of Sci. and Higher Ed., Korea Research
Foundation, Ministry of Sci., Ed. and Sports of the Rep. Of
Croatia, Russian Ministry of Sci. and Tech, and RosAtom of Russia.
\end{scilastnote}

\author{B.~I.~Abelev,$^{8}$ M.~M.~Aggarwal,$^{31}$
Z.~Ahammed,$^{48}$ A.~V.~Alakhverdyants,$^{18}$
I.~Alekseev,$^{16}$ B.~D.~Anderson,$^{19}$
% ...
\\
\normalsize{$^{1}$Argonne National Laboratory, Argonne, Illinois
60439, USA}\\
\normalsize{$^{2}$University of Birmingham, Birmingham, United
Kingdom}\\
\normalsize{$^{3}$Brookhaven National Laboratory, Upton, New York
11973, USA}\\
\normalsize{$^{4}$University of California, Berkeley, California
94720, USA}\\
\normalsize{$^{5}$University of California, Davis, California
95616, USA}\\
\normalsize{$^{6}$University of California, Los Angeles,
California 90095, USA}\\
\normalsize{$^{7}$Universidade Estadual de Campinas, Sao Paulo,
Brazil}\\
\normalsize{$^{8}$University of Illinois at Chicago, Chicago,
Illinois 60607, USA}\\
\normalsize{$^{9}$Creighton University, Omaha, Nebraska 68178,
USA}\\
\normalsize{$^{10}$Czech Technical University in Prague, FNSPE,
Prague, 115 19, Czech Republic}\\
\normalsize{$^{11}$Nuclear Physics Institute AS CR, 250 68
\v{R}e\v{z}/Prague, Czech Republic}\\
\normalsize{$^{12}$University of Frankfurt, Frankfurt, Germany}\\
\normalsize{$^{13}$Institute of Physics, Bhubaneswar 751005,
India}\\
\normalsize{$^{14}$Indian Institute of Technology, Mumbai,
India}\\
\normalsize{$^{15}$Indiana University, Bloomington, Indiana 47408,
USA}\\
\normalsize{$^{16}$Alikhanov Institute for Theoretical and
Experimental Physics, Moscow, Russia}\\
\normalsize{$^{17}$University of Jammu, Jammu 180001, India}\\
\normalsize{$^{18}$Joint Institute for Nuclear Research, Dubna,
141 980, Russia}\\
\normalsize{$^{19}$Kent State University, Kent, Ohio 44242, USA}\\
\normalsize{$^{20}$University of Kentucky, Lexington, Kentucky,
40506-0055, USA}\\
\normalsize{$^{21}$Institute of Modern Physics, Lanzhou, China}\\
\normalsize{$^{22}$Lawrence Berkeley National Laboratory,
Berkeley, California 94720, USA}\\
\normalsize{$^{23}$Massachusetts Institute of Technology,
Cambridge, MA 02139-4307, USA}\\
\normalsize{$^{24}$Max-Planck-Institut f\"ur Physik, Munich,
Germany}\\
\normalsize{$^{25}$Michigan State University, East Lansing,
Michigan 48824, USA}\\
\normalsize{$^{26}$Moscow Engineering Physics Institute, Moscow
Russia}\\
\normalsize{$^{27}$City College of New York, New York City, New
York 10031, USA}\\
\normalsize{$^{28}$NIKHEF and Utrecht University, Amsterdam, The
Netherlands}\\
\normalsize{$^{29}$Ohio State University, Columbus, Ohio 43210,
USA}\\
\normalsize{$^{30}$Old Dominion University, Norfolk, VA, 23529,
USA}\\
\normalsize{$^{31}$Panjab University, Chandigarh 160014, India}\\
\normalsize{$^{32}$Pennsylvania State University, University Park,
Pennsylvania 16802, USA}\\
\normalsize{$^{33}$Institute of High Energy Physics, Protvino,
Russia}\\
\normalsize{$^{34}$Purdue University, West Lafayette, Indiana
47907, USA}\\
\normalsize{$^{35}$Pusan National University, Pusan, Republic of
Korea}\\
\normalsize{$^{36}$University of Rajasthan, Jaipur 302004,
India}\\
\normalsize{$^{37}$Rice University, Houston, Texas 77251, USA}\\
\normalsize{$^{38}$Universidade de Sao Paulo, Sao Paulo, Brazil}\\
\normalsize{$^{39}$University of Science \& Technology of China,
Hefei 230026, China}\\
\normalsize{$^{40}$Shandong University, Jinan, Shandong 250100,
China}\\
\normalsize{$^{41}$Shanghai Institute of Applied Physics, Shanghai
201800, China}\\
\normalsize{$^{42}$SUBATECH, Nantes, France}\\
\normalsize{$^{43}$Texas A\&M University, College Station, Texas
77843, USA}\\
\normalsize{$^{44}$University of Texas, Austin, Texas 78712,
USA}\\
\normalsize{$^{45}$Tsinghua University, Beijing 100084, China}\\
\normalsize{$^{46}$United States Naval Academy, Annapolis, MD
21402, USA}\\
\normalsize{$^{47}$Valparaiso University, Valparaiso, Indiana
46383, USA}\\
\normalsize{$^{48}$Variable Energy Cyclotron Centre, Kolkata
700064, India}\\
\normalsize{$^{49}$Warsaw University of Technology, Warsaw,
Poland}\\
\normalsize{$^{50}$University of Washington, Seattle, Washington
98195, USA}\\
\normalsize{$^{51}$Wayne State University, Detroit, Michigan
48201, USA}\\
\normalsize{$^{52}$Institute of Particle Physics, CCNU (HZNU),
Wuhan 430079, China}\\
\normalsize{$^{53}$Yale University, New Haven, Connecticut 06520,
USA}\\
\normalsize{$^{54}$University of Zagreb, Zagreb, HR-10002,
Croatia}}

\end{document}